\documentclass[12pt,twoside,pra,aps,amssymb,amsfonts,amsmath,tightenlines]{revtex4}

\usepackage{amssymb}
\usepackage{amsmath}
\usepackage{amsfonts}
\usepackage{amsbsy}
\usepackage{amstext}
\usepackage{dsfont}

\usepackage{graphicx}
\usepackage{fancyhdr}
\usepackage{lscape}
\usepackage{pdflscape}

\newtheorem{theorem}{Theorem}

\newtheorem{proposition}[theorem]{Proposition}
\newtheorem{lemma}[theorem]{Lemma}

\newenvironment{proof}[1][Proof.]{\begin{trivlist}
\item[\hskip \labelsep {\bfseries #1}]}{\end{trivlist}}

\newcommand{\re}{\mbox{$\rm e$}}
\newcommand{\rd}{\mbox{$\rm d$}}

\begin{document}

\title{On the Pricing of Storable Commodities}

\author {\textsc{Dorje C.~Brody$^{1,2}$, Lane P.~Hughston$^{3}$, Xun Yang$^{4}$}}
\affiliation{$^{1}$Department of Mathematics, University of Surrey, Guildford GU2 7XH, United Kingdom\\ 
$^{2}$St Petersburg National Research University of Information Technologies, Mechanics and Optics,
49 Kronverksky Avenue, St Petersburg 197101, Russia \\ 
$^{3}$Department of Computing, Goldsmiths University of London, New Cross, London SE14 6NW, United Kingdom \\ 
$^{4}$Shell International Limited, Shell Centre, London SE1 7NA, United Kingdom}

\date{\today}

\begin{abstract}
\noindent This paper introduces an information-based model for the pricing of storable commodities such as crude oil and natural gas. The model uses the concept of market information about future supply and demand as a basis for valuation. Physical ownership of a commodity is taken to provide a stream of convenience dividends equivalent to a continuous cash flow. The market filtration is assumed to be generated jointly by (i) current and past levels of the dividend rate, and (ii) partial information concerning the future of the dividend flow. The price of a commodity is the expectation under a suitable pricing measure of the totality of the discounted risk-adjusted future convenience dividend, conditional on the information provided by the market filtration. In the situation where the dividend rate is modelled by an Ornstein-Uhlenbeck process, the prices of options on commodities can be derived in closed form. The approach that we present can be applied to other assets that yield potentially negative effective cash flows, such as real estate, factories, refineries, mines, and power generating plants.

\vspace{0.05cm}


\noindent \footnotesize{To appear as Chapter 17 in  \emph{Financial Informatics: An Information-Based Approach to Asset Pricing}. D. C. Brody, L. P. Hughston \& A. Macrina (editors). Singapore: World Scientific Publishing Company (2022).} 

\end{abstract}

\maketitle

\section{Introduction}
\label{sec:1}

\noindent In the extensive literature devoted to the pricing and risk management of commodity derivatives, most investigations take as a starting point the specification of a dynamical model for the price process 
for the commodity. The outcome of chance in the market in which the commodity is 
traded is usually represented in such studies with the specification of a fixed probability space equipped with the filtration generated by a Brownian motion of one or more dimensions, and it is typically assumed that the commodity price can be modelled as an Ito process adapted to this filtration. 
Such an approach to the pricing of commodities and related derivatives is in line with the ``standard" 
modelling framework for asset pricing within which much or event most of modern finance theory has been pursued \cite{Karatzas-Shreve, duffie}. 

Nevertheless, there is a fundamental methodological issue in the standard framework: namely,  that the market filtration is fixed in an essentially \textit{ad hoc} way, and that no indication is provided concerning the nature of the information it purports to convey, or why it is relevant to the price. The information is in practice normally no more than that of the 
price movements themselves, so it can hardly be claimed in any useful way that the price movements are taking place ``in response'' to ``shocks'' associated with the arrival of information, for the shocks, as we have said, are no more than the asset price movements themselves. 

One knows, however, that in real markets, information concerning the possible future cash flows and other benefits or obligations linked to the physical possession of an asset can be crucial in the determination of trading decisions, even in situations where such information is imperfect. The movement of the price of an asset should thus be regarded as a \textit{derived} concept, induced by the flow of information to market participants. This is the point of view put forward in the information-based asset pricing theory of Brody, Hughston \& Macrina  \cite{bhm1, bhm2, bhm3} that forms the analytical basis of the present investigation; see also \cite{macrina, rutkowski}.  

The goal of this paper is to incorporate the role played by 
forward-looking information in commodity markets  in such a context, and to derive a model for the
prices used as underlyings in the valuation of commodity derivatives. 
Specifically, we make use of the concept of market information about future supply and 
demand as a basis for the valuation of storable commodities. 
The theory of commodity pricing from a modern perspective has a long history, starting from the work of Black \cite{black1} and Brennan \& Schwartz \cite{bs1}. For a detailed treatment of theory of storage, convenience yield, and related concepts, the reader can 
be referred to Geman \cite{geman1} and references cited therein.
In our approach, we shall assume that the possession of one standard unit of a commodity 
provides a net ``convenience dividend'' equivalent to a cash flow $\{ X_{t} \}_{t \geq 0}$. 
We thus work directly with the actual flow of benefit arising from the possession of the commodity, rather than the percentage convenience yield. The point 
is that the percentage convenience yield so often used in commodity modelling is a secondary notion, since it depends 
on the price, which is what we are trying to determine. In what follows, we present a simple model 
for the convenience dividend process $\{X_{t}\}$. Additionally, we introduce a market information 
process $\{\xi_t\}_{t\geq 0}$ that provides partial or speculative information about the future dividend flow. 
The market filtration is then assumed to be generated jointly by these two processes. In that sense, 
we are explicitly constructing the market filtration in such a way that it contains information relevant 
to the commodity price. Given the market filtration, the price of the commodity is taken to be the 
risk-adjusted discounted expected value of the totality of the future convenience 
dividends. We model $\{X_{t}\}$ by an Ornstein-Uhlenbeck process. We model $\{\xi_t\}$ by a 
process that consists of two terms: a ``signal'' term containing information about the future 
convenience dividend flow, and a ``noise'' term given by an independent Brownian motion. By use 
of this information-based model we are able to derive closed-form expressions for both the price 
of the commodity and for the prices of associated derivatives. 

The remainder of the paper is organized as follows. In \S\ref{sec:2} we introduce our model for the net convenience dividend and for the market filtration. In \S\ref{sec:3} various useful facts 
relating to the Ornstein-Uhlenbeck (OU) process are recalled -- in particular, 
certain features of the OU bridge. These are used 
in \S\ref{sec:4} to show in Proposition 1 that the information process and the convenience dividend rate are jointly Markovian, 
and in \S\ref{sec:5} to derive an expression for the commodity price. Finally,
in \S\ref{sec:7} we present pricing formulae for call options on the underlying 
spot price. 

\section{Information-based commodity pricing}
\label{sec:2}

In the information-based approach of Brody, Hughston \& Macrina \cite{bhm1, bhm2, bhm3}, 
the starting point is the 
specification of (i) a set of one or more random variables (called ``market factors'') 
determining the cash flows associated with a given asset, and (ii) a set of one or more 
random processes (called ``information processes'') determining the flow of information 
to market participants concerning these market factors. The setup, more specifically, 
is as follows. We model the outcome of chance in a commodity market with the specification 
of a probability space 
$({\Omega}, {\mathcal F}, {\mathbb Q})$. The market is not assumed to be 
complete, but we do assume the existence of a preferred pricing measure (or ``risk neutral" measure)
${\mathbb Q}$. In more detail, we assume the existence of a physical measure and a pricing kernel that with respect to the physical measure takes the form of a discount factor times a strictly positive martingale. The martingale is used to change the measure to ${\mathbb Q}$, and from that point onward we formulate the details of the theory with reference to that measure. Then if $\{X_{t}\}$ represents the net convenience dividend, which is given by the benefits associated with possession of the commodity less storage costs and any other direct costs associated with the said possession,
the price of the commodity at time $t$ is given by 
\begin{eqnarray}
S_t = \frac{1}{\pi_t}\, {\mathbb E} 
\left[\left.  \int_t^{\infty} \pi_{u} X_{u} \rd u \right| {\mathcal F}_t \right] , 
\label{asset price}
\end{eqnarray}
where the expectation is taken under ${\mathbb Q}$. Here the discount factor 
$\{\pi_t\}_{t\geq 0}$ is given in terms of the short rate $\{r_t\}_{t\geq 0}$ by 
\begin{eqnarray}
\pi_t = \exp \left (-\int ^{t}_{0} r_{s} \rd s \right).
\label{eq:2}
\end{eqnarray}
The associated money-market account process is then given by $\{1/\pi_t\}_{t\geq 0}$. For simplicity, we assume that the 
default-free interest 
rate system is deterministic.
The market filtration $\{{\mathcal F}_t\}$, with respect to which the conditioning is taken in (\ref{asset price}), is taken  to be generated jointly by (a) the 
convenience dividend process $\{X_t\}_{t \geq 0}$ and (b) a market information 
process $\{\xi_t\}_{t\geq0}$ of the form
\begin{eqnarray}
\xi_{t} = \sigma t \int_t^\infty \pi_u X_u \rd u + B_t \, , 
\label{information process}
\end{eqnarray}
representing partial or noisy information about the future dividend flow. The 
parameter $\sigma$ determines the 
rate at which information about the future dividend stream is revealed to the market. 
The ${\mathbb Q}$-Brownian motion $\{B_t\}$ represents noise arising from rumour, 
baseless speculation, uninformed trading, fake news, and the like, and is assumed to be 
independent of the dividend process $\{X_t\}$. Therefore for each $t\geq 0$ we have 
\begin{eqnarray}
{\mathcal F}_t = {\sigma} \left[ 
\{\xi_s\}_{0 \leq s\leq t},  \{X_s\}_{0 \leq s \leq t} \right].
\end{eqnarray}

The next step is to specify the form of the dividend process. We consider in this paper the case in which $\{X_t\}$ is an Ornstein-Uhlenbeck (OU) 
process. It is worth recalling, by way of 
contrast,  that Gibson \& Schwartz \cite{gibson1, gibson2} assume that the percentage convenience yield should follow a mean-reverting 
process, an approach that has been followed in many subsequent works \cite{schwartz1, hilliard1, hilliard2, schwartz2, schwartz3, miltersen1, schwartz4, casassus1}. The idea of the present work, however, is that a mean-reverting absolute convenience dividend reflects the 
notion that in the long term there is an equilibrium rate of benefit obtained by storing or holding 
the commodity. We thus assume that the dividend process satisfies a stochastic equation 
of the form  
\begin{eqnarray}
\rd X_t = \kappa (\theta - X_t)\rd t + \psi\, \rd \beta_t,
\label{OU process}
\end{eqnarray}
where $\{\beta_t\}$ is a ${\mathbb Q}$-Brownian motion that is independent of 
$\{B_t\}$. We allow for the possibility that the dividend rate may occasionally be negative. The mean reversion level $\theta$, the reversion rate $\kappa$, and 
the dividend volatility $\psi$ are assumed to be constant in the present 
discussion, although the results can be readily generalized to the 
time-dependent case.

\section{Properties of the Ornstein-Uhlenback process}
\label{sec:3}

Before we proceed to work out the conditional expectation (\ref{asset price}), it will be useful to comment on various properties of the Ornstein-Uhlenbeck 
process. These properties, some of which will be well known, but others perhaps less so, 
will help us simplify the calculations. It is an elementary exercise to check that the 
solution to (\ref{OU process}) takes the form  
\begin{eqnarray}
X_t = \re^{-\kappa t} X_0 + \theta (1-\re^{-\kappa t}) + \psi \re^{-\kappa t} 
\int_0^t \re^{\kappa s} \rd \beta_s . 
\label{OU solution}
\end{eqnarray}
The Ornstein-Uhlenbeck process has the property that if we ``reinitialize'' the process 
at time $t$ then its value at some later time $T>t$ can be expressed as
\begin{eqnarray}
X_{T} = \re^{-\kappa (T- t)} X_t + \theta (1-\re^{-\kappa (T - t)}) + \psi 
\re^{-\kappa T} \int_t^T \re^{\kappa u}\, \rd \beta_u.
\label{reinitialize}
\end{eqnarray} 
Since $\{X_t\}$ is a Gaussian process, one can 
verify the following by use of covariance relations: 

\begin{lemma} 
The random variables $X_{t}$ and $X_T-\re^{-\kappa (T-t)}X_t$ 
are independent. 
\label{OU independent components}
\end{lemma}

\noindent \textit{Proof}. The mean of $X_t$ is given by 
\begin{eqnarray}
{\mathbb E}[X_t] = 
\re^{-\kappa t} X_0 + \theta (1-\re^{-\kappa t}). 
\end{eqnarray}	
A straightforward calculation using the Ito isometry then shows that
\begin{eqnarray}
{\rm Var}[X_t] =  \frac{\,\,\psi^2}{2 \kappa}  (1 -   \re^{-2 \kappa t} )
\end{eqnarray} 
and that for $0\leq t\leq T$ we have
\begin{eqnarray}
{\rm Cov}[X_t, X_T] =  \frac{\,\,\psi^2}{\kappa} \re^{- \kappa T} \sinh \kappa t, 
\end{eqnarray} 
from which it follows that
\begin{eqnarray}
{\rm Cov}[X_t, X_T-\re^{-\kappa (T-t)}X_t] = 0,
\end{eqnarray} 
and hence the claimed independence. $\square$
\vspace{.25cm}

\noindent This property of the OU process corresponds to an orthogonal decomposition of the 
form
\begin{eqnarray}
X_T = (X_{T} - \re^{- \kappa (T-t)} X_{t}) + \re^{-\kappa (T-t)} X_{t} 
\label{orthogonal decomposition}
\end{eqnarray}
for $T > t$. The validity of Lemma \ref{OU independent components} can also be checked by direct inspection of  
(\ref{OU solution}) and (\ref{reinitialize}). It should be evident that if the reversion rate is set to zero, then (\ref{orthogonal decomposition}) reduces to the independent-increments decomposition of a 
Brownian motion.

Interestingly, there is another orthogonal decomposition of the OU process that is somewhat 
less obvious than (\ref{orthogonal decomposition}). This is given by the identity 
\begin{eqnarray}
X_t = \left( X_t-\frac{\sinh \kappa t} {\sinh \kappa T} \,
X_{T} \right) + 
\frac{\sinh \kappa t} {\sinh \kappa T} \,
X_{T}. 
\label{OU bridge}
\end{eqnarray}
The process $\{b_{tT}\}_{0 \leq t \leq T}$, defined for fixed $T$ by 
\begin{eqnarray}
b_{tT} = X_t - \frac{\sinh \kappa t} {\sinh \kappa T} 
\, X_{T} ,
\label{OU bridge formula}
\end{eqnarray}
appearing in (\ref{OU bridge}), is an Ornstein-Uhlenbeck (OU) bridge. The OU bridge 
interpolates between the fixed values $b_{0T} = X_0$ and $b_{TT} = 0$, and we are led to another useful result.

\vspace{0.25 cm}
\noindent In particular, a calculation shows the following: 
\begin{lemma} 
The Ornstein-Uhlenbeck bridge $\{b_{tT}\}_{0 \leq t \leq T}$ and the random variable $X_{U}$
are independent  for all $T$ and $U$ such that $0 \leq T \leq U$. 
\end{lemma}

\noindent \textit{Proof}. Since $\{b_{tT}\}_{0 \leq t \leq T}$ and $X_{U}$ are jointly Gaussian, it suffices to show that
for any choice of $t, T, U$ such that $t \leq T \leq U$ the random variables $ b_{tT}$ and $X_{U}$ are independent. We have
\begin{eqnarray}
{\rm Cov}[X_t, X_U] =  \frac{\,\,\psi^2}{\kappa} \re^{- \kappa U} \sinh \kappa t ,
\end{eqnarray} 
and
\begin{eqnarray}
{\rm Cov}[X_T, X_U] =  \frac{\,\,\psi^2}{\kappa} \re^{- \kappa U} \sinh \kappa T ,
\end{eqnarray} 
from which it follows that
\begin{eqnarray}
{\rm Cov}\left[X_t - \frac{\sinh \kappa t} {\sinh \kappa T} 
\, X_{T}, X_U \right] = 0,
\end{eqnarray} 
and hence the claimed independence. $\square$
\vspace{.25cm}

\noindent We note that the mean and variance of the OU bridge are 
given, respectively, by 
\begin{eqnarray}
{\mathbb E}[b_{tT}] = \frac{\sinh \kappa(T-t)}{\sinh \kappa T} X_0 + \left[ 
1 - \frac{\sinh \kappa t+\sinh \kappa(T-t)}{\sinh \kappa T} \right] \theta
\end{eqnarray}
and 
\begin{eqnarray}
\textrm{Var}[b_{tT}] = \frac{\,\,\psi^2}{\kappa} \sinh \kappa t 
\left[ \cosh \kappa t - \frac{\sinh \kappa t}{\sinh \kappa T} \cosh \kappa T \right] .
\end{eqnarray}	

\section{Markov property of market information}
\label{sec:4}

When working with conditional expectations, we often use the shorthand $\mathbb E [Y | Z]$ in place of $\mathbb E [Y |  \sigma \{Z\}]$, where $ \sigma \{Z\}$ denotes the $\sigma$-algebra generated by $Z$. Keeping in mind this notation, we frequently make use of the following. Let $X$, $Y$, and $Z$ be random variables, and assume that $X$ is integrable. Then if $ \sigma \{X, Y\}$ and $ \sigma \{Z\}$ are independent it holds that  
\begin{eqnarray}
\mathbb E [X | \, Y, Z] = \mathbb E [X | Y]. 
\end{eqnarray}
See, for example, Williams \cite{Williams}, section 9.7. We proceed to work out the conditional expectation 
in (\ref{asset price}) to determine the commodity price. The following result will facilitate the 
calculations. 

\begin{proposition} 
The information process $\{\xi_t\}$ and the dividend rate $\{X_t\}$ 
are jointly Markov.
\end{proposition}

\begin{proof} 
We need to show that for $0 \leq t \leq u$ it holds that
\begin{eqnarray}
{\mathbb Q} \left[\left. \xi_u < a \, \cap \,  X_u < b \, \right| \{\xi_s\}_{0\leq s\leq t}, 
\{X_s\}_{0\leq s\leq t} \right] =  {\mathbb Q} \left[\left. \xi_u < a  \, \cap \,    X_u < b \, 
\right| \xi_t, X_t \right].\,\,
\label{Markov property} 
\end{eqnarray}
Let us define a process $\{\eta_t\}$ by setting 
\begin{eqnarray}
\eta_t  = 
\sigma t \int_0^\infty \pi_u X_u \rd u + B_t. 
\end{eqnarray}
It should be evident that 
\begin{eqnarray}
{\sigma}[\{\xi_s\}_{0\leq s\leq t},
\{X_s\}_{0\leq s\leq t}]={\sigma}[\{\eta_s\}_{0\leq s\leq t},
\{X_s\}_{0\leq s\leq t}]. 
\end{eqnarray}
This follows from the fact that
\begin{eqnarray}
\eta_t  = 
\xi_t + \sigma t \int_0^t \pi_u X_u \rd u.
\end{eqnarray}
We observe that $\{\eta_t\}$ is 
Markov in its own filtration. To see this, it suffices to verify that
\begin{eqnarray}
{\mathbb Q}\left( \eta_t\leq x \, \big| \, \eta_s,\eta_{s_1},\eta_{s_2},\ldots,
\eta_{s_k}\right) = {\mathbb Q}\left(\left. \eta_t\leq x\right|\eta_s\right) 
\label{eta Markov property} 
\end{eqnarray}
for any collection of times $t,s,s_1,s_2,\ldots,s_k$ such that $t\geq s\geq s_1
\geq s_2\geq\cdots\geq s_k>0$. Now, it is an elementary property of Brownian 
motion that for any times $t,s, s_1$ satisfying $t>s>s_1>0$ the random variables
$B_t$ and $B_s/s-B_{s_1}/s_1$ are independent. More generally, for $s>s_1>s_2
>s_3>0$, we find that $B_s/s-B_{s_1}/s_1$ and $B_{s_2}/s_2-B_{s_3}/s_3$ are 
independent. Observing  that $\eta_s/s-\eta_{s_1}/s_1 = B_s/s-B_{s_1}/s_1$, 
we find that 
\begin{eqnarray}
{\mathbb Q}\left( \eta_t\leq x \, \big| \, \eta_s,\eta_{s_1}, \ldots,
\eta_{s_k}\right) &=& {\mathbb Q}\left(\eta_t\leq x \, \big| \,
\eta_s, \frac{\eta_s}{s}-\frac{\eta_{s_1}}{s_1}, \cdots,
\frac{\eta_{s_{k-1}}}{s_{k-1}} -
\frac{\eta_{s_k}}{s_k}\right) \nonumber \\ &=& 
{\mathbb Q}\left(\eta_t\leq x \, \big| \, \eta_s, \frac{B_s}{s}-
\frac{B_{s_1}}{s_1}, 
\cdots, \frac{B_{s_{k-1} }}{s_{k-1}}-\frac{B_{s_k}}{s_k}\right). \nonumber \\
\end{eqnarray}
But since $\eta_t$ and $\eta_s$ are jointly independent of $B_s/s-B_{s_1}/s_1$, 
$\cdots$, the Markov property (\ref{eta Markov property}) follows for $\{\eta_t\}$. Let us now define
\begin{eqnarray}
{\mathcal G}_t = {\sigma} \left(\left\{\frac{\eta_t}{t}-\frac{\eta_{s}}{s} 
\right\}_{0 < s \leq t}\right).
\end{eqnarray}
Then clearly we have
\begin{eqnarray}
{\mathcal G}_t = {\sigma} \left(\left\{\frac{B_t}{t}-\frac{B_{s}}{s} 
\right\}_{0 < s \leq t}\right), 
\end{eqnarray}
that is to say, ${\mathcal G}_t$ is generated by the Brownian bridge underlying the noise component of the information process. Note that the sigma algebras $ \sigma [\eta_t$, $\{X_s\} ]$ and ${\mathcal G}_t$ are independent. As a consequence, writing
\begin{eqnarray}
F[a, b \, | \,{\mathcal F}_t] = {\mathbb P} \left[\left. \xi_u < a \, \cap \,  X_u < b \, \right| \,{\mathcal F}_t \right] 
\end{eqnarray}
for the conditional bivariate distribution function, we 
have
\begin{eqnarray}
F[a, b \, | \,{\mathcal F}_t] &=& {\mathbb E} \left[ \mathds 1( \left. \xi_u < a) \mathds 1 ( X_u < b) \, \right| \,{\mathcal F}_t \right] 
 \nonumber \\ 
&=& {\mathbb E} \left[\mathds 1( \left. \xi_u < a) \mathds 1 ( X_u < b) 
\right| \{\xi_s\}_{0 \leq s \leq t}, \{X_{s}\}_{0 \leq s \leq t} \right] \nonumber \\ 
&=& 
{\mathbb E} \left[\mathds 1( \left. \xi_u < a) \mathds 1 ( X_u < b) 
\right| \{\eta_s\}_{0 \leq s \leq t}, \{X_{s}\}_{0 \leq s \leq t} \right] \nonumber \\
&=& {\mathbb E} \left[\mathds 1( \left. \xi_u < a) \mathds 1 ( X_u < b) \right| \eta_t, {\mathcal G}_t, 
\{X_{s}\}_{0 \leq s \leq t} \right] \nonumber \\ 
&=& {\mathbb E} \left[\mathds 1( \left. \xi_u < a) \mathds 1 ( X_u < b) \right| \eta_t, 
\{X_{s}\}_{0 \leq s \leq t} \right] \nonumber \\
&=& {\mathbb E} \left[\mathds 1( \left. \xi_u < a) \mathds 1 ( X_u < b) \right| \xi_t, 
\{X_{s}\}_{0 \leq s \leq t} \right].
\end{eqnarray}
On the other hand, recalling the definition of the OU bridge given by (\ref{OU bridge formula}), we have
\begin{eqnarray}
{\sigma}[\xi_t,\{X_s\}_{0\leq s\leq t}]={\sigma}
[\xi_t, X_t, \{b_{st}\}_{0\leq s\leq t}]. 
\end{eqnarray}
Now, it is easy to see that  $\{b_{st}\}_{0\leq s\leq t}$ 
and $\{X_u\}_{u \geq t}$ are independent. It follows that $\sigma [\{b_{st} \}_{0\leq s\leq t} ]$
and $\sigma [\xi_t, X_t, \xi_u, X_u]$ are independent, from which we get (\ref{Markov property}). 
\end{proof}

\section{Commodity pricing formula}
\label{sec:5}

The joint Markov property (\ref{Markov property}) implies that 
\begin{eqnarray}
{\mathbb E} \left[\left. \int_t^\infty \pi_u X_u  \, \rd u \, \right| \{\xi_s\}_{0\leq s\leq t}, 
\{X_s\}_{0\leq s\leq t} \right] =  {\mathbb E} \left[\left. \int_t^\infty \pi_u X_u \, \rd u \,
\right| \xi_t, X_t \right],
\end{eqnarray}
which allows one to reduce the problem of working out 
the commodity price  (\ref{asset price}) to that of calculating 
\begin{eqnarray}
S_t = \frac{1}{\pi_{t}}\, {\mathbb E} \left[\left. \int_t^\infty \pi_u X_u \, \rd u  \, \right| 
\xi_t, X_t \right]. 
\label{asset price reduced} 
\end{eqnarray}
One observes that from the orthogonal decomposition (\ref{orthogonal decomposition}) we can
isolate the dependence of the commodity price on the current level of the convenience 
dividend rate $X_{t}$. Remarkably, the dependence turns out to be linear. That is, 
we have
\begin{eqnarray}
\int_t^\infty \pi_u X_u \, \rd u = \int_t^\infty \pi_u \left( X_u - \re^{-\kappa (u-t)}X_t 
\right) \rd u + \left( \int_t^\infty \pi_{u} \re^{-\kappa (u-t)}\, \rd u \right) X_{t} . \,\, 
\end{eqnarray}
Substituting this formula in equation (\ref{asset price reduced}), we deduce that 
\begin{eqnarray}
\pi_t S_t &=& {\mathbb E} \left[ \left. \int_t^\infty \pi_u 
\left(X_u - \re^{-\kappa (u-t)} 
X_t \right) \rd u \right| \xi_t, X_t \right ] \nonumber \\ && \qquad + {\mathbb E} 
\left[\left.  \left( \int_t^\infty \pi_u \re^{-\kappa(u-t)} \,  \rd u \right) X_t \right| \xi_t, 
X_t \right ] \nonumber \\ &=& 
{\mathbb E} \left[\left. A_t \,  \right| \xi_t, 
X_t \right ] + \left( \int_t^\infty 
\pi_u \re^{-\kappa(u-t)} \, \rd u \right) X_t ,
\end{eqnarray}
where
\begin{eqnarray} 
A_t = \int_t^\infty \pi_u (X_u - \re^{-\kappa(u-t)}X_t) \rd u.
\label{random variable A} 
\end{eqnarray}
Next we observe that as a consequence of (\ref{information process}) and (\ref{random variable A})  we have
\begin{eqnarray} 
\xi_t = \sigma t \left [ A_t + \left(  \int_t^\infty 
\pi_u \re^{-\kappa(u-t)} \rd u \right) X_t \right ] + B_t .
\end{eqnarray}
It follows that 
\begin{eqnarray}
\pi_t S_t  =
{\mathbb E} \left[\left. A_t \,  \right| \sigma t A_t + B_t, X_t \right ] + \left( \int_t^\infty 
\pi_u \re^{-\kappa(u-t)} \rd u \right) X_t .
\end{eqnarray}
Note that the conditioning with respect to 
$X_t$ in the first term above drops out since by Lemma (\ref{OU independent components}) the random variables 
$\{X_{u} - \re^{-\kappa (u-t)}X_{t}\}_{u\geq t}$ and $X_t$ are independent,  which allows one to deduce that the sigma-algebras $\sigma \{X_t\}$ and $\sigma \{A_t, B_t\}$ are independent. Therefore,
\begin{eqnarray}
\pi_t  S_t =
{\mathbb E} \left[\left. A_t \,  \right| \sigma t A_t + B_t \right ] + \left( \int_t^\infty 
\pi_u \re^{-\kappa(u-t)} \rd u \right) X_t .
\end{eqnarray}
The problem of determining the commodity price is thus reduced to that of calculating a conditional 
expectation of the form ${\mathbb E}[A_t|A_t+C_t]$ for $t>0$, where $A_t$ is given 
by (\ref{random variable A}) and $C_t=B_t/\sigma t$. We observe that $A_t$ and $C_t$ are 
independent Gaussian random variables. To compute the 
conditional expectation above, we recall another result concerning orthogonal 
decompositions of Gaussian random variables: 
\begin{lemma}
If $A$ and $C$ are 
independent Gaussian random variables, then  $A+C$ and 
$(1-z)A-zC$ are independent if $z={\rm Var}[A]/
({\rm Var}[A]+{\rm Var}[C])$.  
\end{lemma}
\noindent In view of this observation, let us express $A_t$ in the form 
\begin{eqnarray}
A_t = z_t(A_t+C_t) + (1-z_t)A_t - z_t C_t, 
\end{eqnarray}
where 
\begin{eqnarray}
z_t = \frac{{\rm Var}[A_t]}{{\rm Var}[A_t]+{\rm Var}[C_t]}  . 
\end{eqnarray}
Then we find that
\begin{eqnarray}
{\mathbb E}[A_t |A_t+C_t] = z_t(A_t+C_t) + (1-z_t){\mathbb E}[A_t] - z_t {\mathbb E}[C_t] . 
\end{eqnarray} 
Clearly, we have ${\mathbb E}[C_t]=0$. Furthermore, if we set $T=u$ in equation (\ref{reinitialize}) 
we deduce that 
\begin{eqnarray}
{\mathbb E}[A_t] &=& {\mathbb E}\left[ \theta \int_t^\infty \pi_{u} 
\left(1-\re^{-\kappa (u - t)}\right) \rd u + \psi \int_t^\infty \re^{-\kappa u}\pi_{u} 
\int^{u}_{t} \re^{\kappa s} \rd \beta_{s} \rd u \right ] \nonumber \\ &=& 
\theta \int_t^\infty \pi_u \rd u - \theta\int_t^\infty \pi_u \re^{-\kappa(u - t)} \rd u . 
\end{eqnarray}
The final step in deriving the commodity price is to determine the variances of $A_t$ and $C_t$. 
To simplify the notation let us write 
\begin{eqnarray}
p_{t}=\int_t^\infty \pi_u \rd u, \qquad  
q_t = \int_t^\infty \pi_u \re^{-\kappa (u-t)} \rd u. 
\end{eqnarray}
Then a short calculation shows that 
\begin{eqnarray}
{\rm Var}[A_t] = \psi^2 \int_t^\infty q_s^2 \rd s, \qquad 
{\rm Var}[C_t] = \frac{1}{\sigma^{2} t} ,
\end{eqnarray}
and hence that  
\begin{eqnarray}
z_t = \frac{\sigma^{2}\psi^{2}t\int_t^\infty q^{2}_{s}\rd s}
{1+ \sigma^{2}\psi^{2}t\int_t^\infty q^{2}_{s}\rd s } . 
\label{z}
\end{eqnarray}
Putting these results together, we deduce that the price of the commodity at 
$t$ is given by 
\begin{eqnarray}
\pi_t S_t = (1-z_t) \, \left [ \theta p_{t} + q_{t}(X_{t}- \theta)\right ] 
+  \frac{1}{\sigma t} z_t  \, \xi_{t} . 
\label{commodity price}
\end{eqnarray}
Observe that the first term in (\ref{commodity price}) is essentially the annuity valuation of a 
constant dividend rate set at the reversion level $\theta$, together with a correction 
term to adjust for the present level of the dividend rate. The second term, on the 
other hand, represents the contribution from the noisy observation of the future 
dividend flow. 

Several interesting observations can be made regarding the weight factor (\ref{z}), which lies between zero and one for all $t$. For large 
$\psi$ and/or large $\sigma$, the value of $z_t$ 
tends to unity; for small $\psi$ and/or small $\sigma$, the value of $z$ 
tends to 0. Hence, if the market information has a low noise content, or if the volatility 
of the convenience dividend is high, then market participants also rely heavily on the 
information available about the future in their determination of the price, rather than 
assuming that the current value of the dividend is a good indicator for the future. 

Conversely, in the absence of a strong signal concerning the future dividend flow, 
an annuity valuation based on the current dividend level will dominate the price. We 
see therefore that important intuitive characteristics are encoded explicitly in the 
pricing formula (\ref{commodity price}). Indeed, (\ref{commodity price}) captures rather well the idea of 
information-based asset pricing, showing how varying amounts of information about 
the future can affect the development of prices, and that prices typically represent a 
kind of compromise between what we know for sure at some given time, and the 
less trustworthy but nevertheless significant intelligence that we may possess regarding 
events that lay ahead. 

In the special case for which the interest rate is constant, the valuation formula 
(\ref{commodity price}) simplifies somewhat to give the following: 
\begin{eqnarray}
S_t = (1-z_{t}) \frac{1}{r} \left[\frac{\kappa}{r+\kappa}\theta + 
\frac{r}{r+\kappa}X_{t}\right] + \frac{\re^{rt}}{\sigma t}\, z_{t}\, \xi_t, 
\label{constant parameter price}
\end{eqnarray}
where the weight factors are
\begin{eqnarray}
z_t = \frac{\sigma^{2}\psi^{2}t}{2r\left(r+\kappa\right)^{2}\re^{2rt}
+\sigma^{2}\psi^{2}t}, \quad 1-z_t= \frac{2r\left(r+\kappa\right)^{2}\re^{2rt}}
{2r\left(r+\kappa\right)^{2}\re^{2rt}+\sigma^{2}\psi^{2}t}. 
\label{eq:28}
\end{eqnarray}

We have performed Monte Carlo simulation studies to gain further intuition 
concerning the dynamical behaviour of the commodity price. Furthermore, we have 
calibrated the model parameters to the prices of crude oil, and we have compared 
the resulting simulated sample paths to market data. In the case of the crude oil 
markets we are able to estimate the expected long-term future spot price from the historical 
average of spot prices, since there exists a supply-demand equilibrium price level 
to which the long-run price tends to converge. The results indicate that even in the 
constant-parameter model considered above, the model is sufficiently rich to capture 
elements of the behaviour of market data 

\section{Pricing commodity derivatives}
\label{sec:7}

We now return to the price process (\ref{constant parameter price}) in the case of a constant interest rate and work out  the value of a European-style call option with strike $K$ and maturity $T$.  Since $\mathbb Q$ is the pricing measure, we have
\begin{eqnarray}
C_{0} = \re^{-rT}\mathbb{E}\left[\left(S_{T} - K\right)^{+}\right] 
\label{option valuation formula}.
\end{eqnarray} 
We observe that $S_T$ consists of a linear combination of
three random components, namely  $X_T$, $\int_T^\infty\re^{-ru}X_{u}\rd u$, and $B_T$, and 
that all three components are Gaussian. It follows that $S_T$ is also Gaussian, and therefore we can write
\begin{eqnarray}
C_0 = \re^{-rT}\frac{1}{\sqrt{2\pi {\rm Var}[S_T]}} 
\int_K^\infty (z-K) \exp\left(-\frac{\left(z-{\mathbb E}[S_T]\right)^{2}}
{2\,{\rm Var}[S_T]} \right)\rd z .  
\label{option price}
\end{eqnarray}
Performing this  integral, we obtain 
\begin{align}
C_0 = \re^{-rT} \left[\sqrt{\frac{{\rm Var}[S_T]}{2\pi}}\exp\left( 
-\frac{({\mathbb E}[S_T]-K)^2}{2{\rm Var}[S_T]}\right) + ({\mathbb E}[S_T]-K) 
N\left(\frac{{\mathbb E}[S_T]-K}{\sqrt{{\rm Var}[S_T]}}\right)\right],
\label{explicit option price}
\end{align}
where $N(x)$ is the normal distribution function. Thus, the 
problem reduces to a determination of the mean and the 
variance of $S_T$. A 
calculation gives
\begin{eqnarray}
{\mathbb E}[S_T] = \frac{1}{r}\left[\frac{\kappa}{r+\kappa}\theta + \frac{r}{r+\kappa} 
\left[\re^{-\kappa T} X_{0} + \theta (1 - \re^{-\kappa T}) \right]\right]
\label{mean of option}
\end{eqnarray}
and 
\begin{eqnarray}
{\rm Var}[S_T] = 
\frac{\psi^{2}}{2\kappa \left(r+\kappa\right)^2}(1-\re^{-2\kappa T}) + z_T^2 
\left[\frac{\psi^{2}}{2r(r+\kappa)^2} + \frac{\re^{2rT}}{\sigma^{2} T}\right].
\label{variance of option}
\end{eqnarray}
Substitution of (\ref{mean of option}) and (\ref{variance of option}) in (\ref{explicit option price}) then gives the option 
price. 
One can also work out the price processes for options. Similar calculations can be carried out to obtain the prices of futures contracts and futures options. 

Although very simple in its structure, the model we have presented captures nicely certain aspects of the pricing of commodities in an information-theoretic framework. Looking ahead, it would be interesting to pursue a similar line of argument in more complex settings, such as those presented by electricity markets. To get a sense of what is involved in that case see \cite {Cartea-Figueroa}. For an alternative approach to the pricing of commodities in an information-based setting see \cite{Macrina-Sekine}. For further discussion of the ideas developed in the present paper see \cite{Yang}. 

\begin{acknowledgments}
\noindent The authors are grateful  to participants at the AMaMeF (Advanced Mathematical Methods in Finance) conferences in Alesund, Norway (2009) and Bled, Slovenia (2010), the Workshop on Derivatives Pricing and Risk Management at the Fields Institute, Toronto (2010), and meetings at the University at Warwick (2011) and the London School of Economics (2012), where parts of this work were presented, for helpful comments. We thank the referee for useful suggestions. XY acknowledges support from Shell International Ltd. 
\end{acknowledgments}

\end{document}